\documentclass[conference]{IEEEtran}
\pdfoutput=1
\usepackage{cite}
\usepackage{amsmath,amssymb,amsfonts}
\usepackage{algorithmic}
\usepackage{graphicx}
\usepackage{textcomp}
\usepackage{xcolor}
\def\BibTeX{{\rm B\kern-.05em{\sc i\kern-.025em b}\kern-.08em
    T\kern-.1667em\lower.7ex\hbox{E}\kern-.125emX}}
\begin{document}

\title{Fairness in Socio-technical Systems: \\a Case Study of Wikipedia
}

\author{\IEEEauthorblockN{Mir Saeed Damadi}
\IEEEauthorblockA{\textit{Département d'Informatique et d'Ingénierie} \\
\textit{Université du Québec en Outaouais}\\
Gatineau, Canada \\
damm15@uqo.ca}

\and
\IEEEauthorblockN{Alan Davoust}
\IEEEauthorblockA{\textit{Département d'Informatique et d'Ingénierie} \\
\textit{Université du Québec en Outaouais}\\
Gatineau, Canada \\
alan.davoust@uqo.ca}

}

\maketitle

\begin{abstract}
Problems broadly known as algorithmic bias frequently occur in the context of complex socio-technical systems (STS), where observed biases may not be directly attributable to a single automated decision algorithm. As a first investigation of fairness in STS, we focus on the case of Wikipedia. We systematically review 75 papers describing different types of bias in Wikipedia, which we classify and relate to established notions of harm from algorithmic fairness research. By analysing causal relationships between the observed phenomena, we demonstrate the complexity of the socio-technical processes causing harm. Finally, we identify the normative expectations of fairness associated with the different problems and discuss the applicability of existing criteria proposed for machine learning-driven decision systems.
\end{abstract}


\section{Introduction}

Fairness has emerged as an important concern in algorithmic systems, mainly focused on the potentially unfair outcomes of automated decisions powered by machine learning~\cite{fairness-testing-chen2022fairness,tab-10-chouldechova2017fair}. Problematic systems have been studied in a variety of domains, including justice~\cite{angwin2016machine}, health care~\cite{ML-03-examples-de2018clinically,ML-04-examples-kourou2015machine} and education~\cite{ML-05-examples-oneto2020learning,ML-06-examples-oneto2017dropout}, and many normative fairness concepts~\cite{binns2018fairness,mehrabi2021survey}, tests~\cite{fairness-testing-chen2022fairness} and corrective measures \cite{tab-demographic1-zafar2017training} have been proposed. 

In this paper, we investigate how fairness concepts may be studied in complex socio-technical systems (STS), possibly including (but not limited to) algorithmic components. In other words, quoting Judith Simon \emph{el al.}~\cite{simon2020algorithmic}, we aim to widen the study of fairness to ``the broader socio-technical system in which technologies are situated". 

In this work we consider STS of the computational kind, or \emph{social machines}~\cite{shadbolt2019theory}, i.e. networks of people interacting over a large-scale digital infrastructure, including various automated and semi-automated components\footnote{Note that we do not claim this to be a universal definition of social machines: we only mean to characterize the types of STS that we have in mind}. The main motivation in analyzing such systems is not to evaluate the risk of using a particular algorithm (e.g. a deep neural network, with its limitations on explainability), but to understand the biases and harms that arise from the interactions of people and algorithms, where the main source of complexity is in the interactions rather than in any specific algorithm.

This will also help us understand how harm might occurs in a socio-technical system where there is no obvious automated component to blame, even in combination with other factors.


 An example of such a situation is the case of abuse in online forums, which disproportionately affects women and minorities. While the behavior of specific participants may be the central problem, the interactions between perpetrators and victims are mediated and shaped by a digital infrastructure, and the lack (or insufficiency) of safeguards within the socio-technical system -- including both the social and the technical level -- can also be blamed for these problematic outcomes. 

An obvious question is how such an analysis might differ from studying the fairness properties of a more narrowly defined algorithmic system or component. 

Working to elucidate this general question, in this paper we expose a case study where we analyse the fairness properties of Wikipedia. Wikipedia is an emblematic STS, with complex interaction processes governed by numerous rules and norms \cite{kittur2008crowdsourcing}, with many automated components involved (including thousands of bots \cite{bot-roles1-zheng2019roles,bot-roles2-geiger2018lives} responsible for 20\% of edits \cite{bot-livingstone2016population}), but no prominent automated decisions that directly affect people as in the usual model of fairness in algorithmic systems. 

Wikipedia being a well-studied system, many types of ``bias" relevant to our problem have already been studied on Wikipedia. We therefore proceed by first systematically reviewing studies of ``bias" in Wikipedia, identifying and categorizing the problematic phenomena discussed in the papers. We then analyze these bias concepts in relation to established notions of \emph{harm} and \emph{fairness} defined for algorithmic systems, and analyze the causal relationships between these phenomena. 

Our systematic review covers 75 papers investigating ``bias" in Wikipedia, and our analysis addresses the following research questions: \begin{itemize}
    \item \textbf{(RQ1)} When fairness-related concepts (such as ``bias") are discussed in the context of Wikipedia, what specific phenomena are being pointed at?
    \item \textbf{(RQ2)} In what sense are these biases considered harmful, and can we relate them back to normative expectations expressed as technical fairness criteria?
    \item \textbf{(RQ3)} Are there any clear causal relationships between these problematic phenomena?
    
\end{itemize}

Our main conclusions are that fairness in STS can meaningfully be analyzed using concepts and techniques developed for algorithmic systems, and secondly that the complexity of the causal relationships between different bias phenomena justifies having a holistic view of the STS, because interventions on any bias phenomena are likely to have consequences on multiple others.

The rest of the paper is organized as follows: in section \ref{sec:review} we present our systematic review of ``bias" in Wikipedia, identifying and categorizing the bias phenomena, then in section \ref{sec:harm} we analyze these phenomena in terms of harm and fairness, before analyzing the causal relationships between the phenomena in section \ref{sec:causal}. We draw our conclusions in section \ref{sec:conclusion}.

\section{Fairness issues in Wikipedia}
\label{sec:review}
We now turn to our first research question:

\textbf{(RQ1)} When fairness-related concepts (such as ``bias") are discussed in the context of Wikipedia, what specific phenomena are being pointed at?

In this section, we begin to answer this question by listing and categorizing the phenomena described as bias in Wikipedia.

\subsection{Methodology}
In order to answer this question, we conducted a systematic review of papers studying fairness (and related concepts) in Wikipedia. In order to identify relevant papers, we searched the Scopus database and Google Scholar using the keywords ``bias", ``fairness", ``discrimination", ``fair" and ``unfair" in conjunction with ``Wikipedia", and collected papers published up to December 2022. In Scopus we limited the scope of the search to titles, abstracts and introductions. We extracted 131 papers from the Scopus database and 30 papers from Google Scholar. In addition to this search targeting fairness-related concepts, we also collected more general surveys of scholarly research about Wikipedia -- using the keywords ``review" or ``survey" and ``Wikipedia" in Google Scholar -- and searched these surveys for other relevant papers and terms to describe fairness-related issues. The terms ``disparity" and ``inequality" were identified in this manner.

The most common fairness-related keyword in the surveyed papers is ``bias", and we discarded papers where the term did not refer to a lack of fairness (in any of the senses discussed previously). In particular, we eliminated papers referring to statistical or cognitive biases, which are not related to fairness.

In several papers we found that ``bias" was used to describe text that didn't follow Wikipedia's \emph{neutral point of view} policy (NPOV): this policy states that articles should ``[represent] fairly, proportionately, and, as far as possible, without editorial bias, all the significant views that have been published by reliable sources on a topic". 
 
 A violation of this policy could include a lack of representation of ``significant views" held by a particular social group (and therefore constitute harm towards the group), or ``editorial bias" portraying certain individuals or social groups in a negative light. We retained papers describing such phenomena, and eliminated the others. Eliminated papers typically discussed automated techniques to identify ``biased" text, with no reference to any particular person or group being potentially affected by the bias.

 We ended up with 75 papers, which almost all described biases against identifiable social groups. One paper discussed the unfair treatment of a single person, in the sense of individual fairness.
 
 In the following, we list the phenomena described as problematic in the papers and categorize them according to the social groups affected by the bias (the case relating to individual fairness being presented separately), and according to the type of harm being described, using the taxonomy proposed by Shelby et al. \cite{harm-shelby2022sociotechnical}. 

\subsection{Social groups affected by bias}

The social groups affected by bias in Wikipedia can be described by three separate attributes (or sets of attributes). 

The first of these attributes is \emph{gender}: numerous papers describe the ``gender gap", ``gender biases" or ``gender disparities". These papers describe several phenomena where women are disadvantaged compared to men. We note we did not find any research discussing biases affecting other gender categories (e.g. non-binary people). 

A second relevant demographic attribute is \emph{culture}, which we will associate here with \emph{language}, \emph{country}, and \emph{race}. The papers that we group in this category describe bias as ``cultural bias", ``geographical bias", ``language-specific bias", ``local bias", ``ingroup bias", the ``colonization of Wikipedia", ``political bias", and other closely related phrases. In these papers, the social groups affected by biases were directly or indirectly described as cultural groups -- residents of certain countries or speakers of certain languages, with the implications that these attributes define relatively homogeneous cultural groups. Regarding racial bias, only one paper was found explicitly reporting racial bias in Wikipedia \cite{adams2019counts}, describing the under-representation of non-white academics in Wikipedia biographies. While we don't want to minimize the importance of the reported bias, it should be noted that the under-representation of the ``Global South" in Wikipedia (which is also described as ``geographical bias") also implies a racial bias, and therefore it makes sense to us to group these group-defining attributes together.

Finally, a small number of papers identify potential bias against the social group of \emph{anonymous Wikipedia contributors}, as opposed to \emph{registered contributors}. This group may have some overlap with the other categories, but this cannot be confirmed as we don't know the demographic make-up of anonymous users. In addition, the potential bias described for this group is different from those described for other demographic groups, and should be analyzed separately.

In the following subsection, we describe for each group the phenomena affecting these groups that have been described as bias in the papers we reviewed. We end with a short subsection discussing the case related to \emph{individual fairness}.

\subsection{Gender bias }

\subsubsection{Under-participation of women} 
In many of the reviewed papers, the terms \textit{gender bias/ gender gap} refer to the disparity between number of male and female contributors (readers, writers, and editors)\cite{03-ratkovic2021women,05-yanisky2015gender,11-glott2010analysis,13-okoli2014wikipedia,16-nielsen2014wikipedia,19-young2020gender}. According to surveys by the Wikimedia Foundation and other analyses, in 2013, the reported figures for female participation range between 13\% and 22\% \cite{06-zandpour2020gender}. While the lack of participation by women -- or any other group -- may not correspond \emph{per se} to any definition of harm, it is more of a symptom of other problems (phenomena reducing women's participation), and it may participate in causing other phenomena that do match definitions of harm.

\subsubsection{Under-representation of Women} 
Two phenomena can be categorized as the under-representation of women.

\paragraph{Wikipedia biographies} Many papers compare the number of biographies of men and women \cite{01-wagner2015s,02-tripodi2021ms,06-zandpour2020gender,19-young2020gender,20-worku2020exploring,21-young2016s}, and find that women are significantly under-represented: in July 2019, only 17.9\% of biographies in Wikipedia were women's biographies \cite{02-tripodi2021ms}. However, it's worth noting that according to Wagner et al.\cite{01-wagner2015s}, a comparison with external databases of so-called \emph{notable people} shows that among these notable people women are not under-represented, in the sense that a notable man and a notable woman (according to the reference datasets for notability used in the paper) have similar chances of having a Wikipedia biography. On the other hand, Adams et al.\cite{adams2019counts} study Wikipedia biographies of American Sociologists, and using bibliometric definitions of notability, they find that for similar notability levels, men are at least twice as likely to have a Wikipedia biography. In other words, there is a clear under-representation of women when compared with the approximate ratio of women to men in the world population in general, but the results are less conclusive when we attempt to control for the (ill-defined) notion of ``notability". This suggests that the bias in not the process of assigning Wikipedia pages to notable people, but rather in the process of becoming ``notable" to begin with.

\paragraph{Deletion of biographies} A related but separate phenomenon is the disparate rate of deletion for biographies of men and women: for one of Wikipedia's page deletion processes (Article for deletion, AFD), a study found that 25\% of the biographies being deleted were women's biographies \cite{02-tripodi2021ms}. While 25\% is much lower than 50\%, the relevant comparison here is with the proportion of women's biographies that do exist and could potentially be deleted (i.e. 17.9\%). According to this analysis, women are therefore over-represented in the biographies deletion process, compared to the overall number of woman's biographies on Wikipedia. 

\paragraph{Structural bias} There is another group of papers that examine the number of links between biographies, according to the genders of the biographies connected by links \cite{01-wagner2015s,06-zandpour2020gender}. These analyses found that the statistical number of links from women's biographies to men's biographies is greater than the number of links from men's biographies to women's biographies. This type of bias is called structural bias in some articles.


\subsubsection{Biased Representation of Women}
In addition to being \emph{under-}represented, women tend to be represented in a biased way on Wikipedia.
\paragraph{Lexical bias}
The lexical content of articles in general, including the particular case of biographies is the primary mechanism of biased representation of women \cite{01-wagner2015s,03-ratkovic2021women,05-yanisky2015gender,06-zandpour2020gender,09-rawatethical,12-mola2012wikipedia,14-mesgari2015sum,15-jullien2012we,20-worku2020exploring}. This type of bias is called lexical bias in some papers \cite{01-wagner2015s}. Lexical bias is visible when the content of men and women's biographies differs following stereotypical patterns: women's biographies contain more information related to their families, their spouse and children, whereas men's biographies contain more information about themselves, their interests, their activities and achievements \cite{01-wagner2015s,06-zandpour2020gender,Event-sun2021men}.

\paragraph{Categorization bias}
Finally, bias based on categorization, which was proposed by Yanisky-Ravid et al. \cite{05-yanisky2015gender}, refers to bias that treats the minority group as an exception within the majority group. For example, Amanda Filipacchi, an American contemporary novelist who has published four novels, she realized in 2013 that she had been re-categorized from the category of ``American novelists" to the category of ``American female novelists". Filipacchi noticed that Wikipedia editors were systematically relegating women (and individuals of ethnic minorities) from the category ``American novelist" to a separate category specifying their special ``other" characteristics. Filipacchi commented on this recategorization in a New York Times article, stating the fact that there is no ``male American novelist" category \cite{05-yanisky2015gender}. We note that Young et al. \cite{19-young2020gender} do not see categorization bias as problematic, but rather as an attempt by some editors to reduce the bias created by other editors. They state: ``while an article on a male CEO may be categorized under CEO and Business, one on a female CEO will likely receive more categories such as CEO, Female CEO, Business, and Women in Business", suggesting a higher visibility and therefore benefits for them.

\subsubsection{Under-representation of women's interests}
A small number of articles have explored whether topics of interest to women are under-represented in Wikipedia \cite{03-ratkovic2021women,05-yanisky2015gender,20-worku2020exploring} That is, researchers compare the number of articles that contain content that is of interest to men with the number of articles that contain content that is of interest to women. Different experiments assigned topics to men and women either using ``well known" stereotypical associations, or by extracting these topics from corpora of magazines intended for male / female audiences.
Based on Worku et al.'s paper \cite{20-worku2020exploring}, articles of presumed interest to women had between 3.2\% to 4.8\% rate of being nominated for deletion by AFD, while articles presumed to be of interest to men ranged from 3.0\% to 3.6\% rate. Regarding speedy deletion (CSD), Articles of supposed interest to women nominated for deletion ranged from 7.6\% to 11.2\%, while articles of supposed interest to men ranged from 6.9\% to 8.1\%. 

\subsection{Cultural and geographical bias}
Bias based on culture and geography, which has been investigated less than gender bias in papers, examines various phenomena affecting cultural groups, broadly defined either by their shared country(ies) of origin or by their languages.

It is remarkable to see that very similar bias phenomena have been reported for cultural groups as for women; and we can conjecture that those phenomena that have \emph{not} been reported for cultural groups could also be present, but more difficult to measure.

\subsubsection{Under-participation}
A similar phenomenon to the under-participation of women can be described here, as a \emph{geographical bias}: a Wikimedia Foundation report \cite{Recial-01-graham2015digital} notes that 45\% of Wikipedia contributors are from only five countries (Italy, France, Great Britain, Germany and the United States). If a single language edition of Wikipedia is considered, a similar situation will occur: the contributors of that edition will be predominantly from the country or countries where the language is spoken. In addition, the demographic make-up of Wikipedia contributors within a given country may not reflect the overall demographic distribution of the country, meaning that there are potentially other ways of defining under-participating groups (e.g. along racial lines).

\subsubsection{Under-representation}
\paragraph{Biographies} 
Following the observation that Wikipedia contributors are mostly from a small number of countries, Beytía \cite{22-beytia2020positioning} points out that 62\% of Wikipedia biographies are also those of people from the same five countries. Considering the social group defined by race in a given country, Adams et al. \cite{adams2019counts} also find that non-white scholars were under-represented in Wikipedia, compared to their white counterparts of similar notability (as measured by bibliometrics).

\paragraph{Cultural bias}
An important aspect of cultural bias is the under-representation of certain cultural groups' perspectives in Wikipedia articles. For example, Young et al. \cite{19-young2020gender} evoke the \emph{colonization of Wikipedia}, meaning the dominance of western perspectives, even on topics primarily associated with other cultures (e.g. traditional Chinese medicine). They point to an example of an edit war over an article describing a plant used in Chinese traditional medicine, where the page's positive bias on the influence of Chinese medicinal plants became negative bias and all its references except the one referring to a western scientific article were removed. Several authors have also studied \emph{linguistic points of view}\cite{alvarez2020linguistic,callahan2011cultural,rogers2012neutral,massa2012manypedia}, i.e. differences in how a topic is described in different Wikipedia language editions. A striking example is how historical international conflicts are portrayed by cultural groups representing the two sides of the conflict, and editing different Wikipedia language editions \cite{alvarez2020linguistic}. 

\subsection{Anonymous users}

In addition to the main social groups discussed previously, for which biases are commonly discussed in the context of algorithmic fairness, a small number of papers \cite{edit-flagging1-teblunthuis2021effects,edit-flagging2de2016profiling,edit-flagging3de2015use} have discussed the risk of bias against the social group of \emph{anonymous editors}, a group that is defined by their status with respect to a technical feature of Wikipedia, namely registration. In this case the form of bias potentially affecting the users is what we might call \emph{over-policing}. Wikipedia allows both registered and anonymous contributors, but the contributions of anonymous editors are likely more closely scrutinized than those of registered users, and more likely to be reverted \cite{edit-flagging1-teblunthuis2021effects}. This is due to semi-automated ``flagging" software, which uses machine learning to detect potentially problematic edits, and flag them for review by ``patrolling" core editors. The interface of the system indicates both a risk score (calculated by the tool) and the registration status of the editor. De Laat \cite{edit-flagging2de2016profiling} suggests that the automated component may not be biased, but the fact of showing the registration status leads patrolling editors to focus on the anonymous users, thus ``profiling" them.

\subsection{Bias against individuals}
Finally, we found that a single paper by Brian Martin \cite{martin2018persistent}, an Australian sociologist, describes bias in a way that relates to the concept of individual fairness. Martin has a Wikipedia biography, and claims that his biography unfairly portrays him in a negative light. His argument implies that his expectation for fairness is defined by individual fairness: he compares his own biography with those of other Australian sociologists of similar age and similar academic status, and observes that his stands out by the negative tone and the lack of recognition of his academic achievements, and instead focusing on reputationally damaging controversies.


\section{Bias phenomena, harm, and fairness}
\label{sec:harm}
\begin{table*}
\begin{center}
\caption{The number of papers describing the different types of bias. The total may exceed 75 as several papers discuss multiple bias issues.}
  \label{tab:paper-list2}
\begin{tabular}{|c|c|c|c|c|} 
 \hline & Gender & Culture and Geography & Anonymous Users & Individual \\ [0.5ex] 
 \hline
under-participation  & 17 & 5 & - & -\\ \hline
under-representation  & 16 & 28 & - & -\\ \hline
biased deletion processes & 11 & - & - & -\\ \hline
biased representation  & 27 & 6 & - & 1\\ \hline
over-policing  & - & - & 3 & -\\ \hline
\end{tabular}
\end{center}
\end{table*}

The above description of biases identified in our systematic review is summarized in table \ref{tab:paper-list2}. The lines in the table categorize the different observed phenomena into a small number of general types of bias, and the columns indicate the groups which are affected by these biases; the last column indicates the case where a single individual is targeted by a form of bias.

We now turn to our next research question, which is about a normative analysis of these biases:

\textbf{(RQ2)} In what sense are these biases considered harmful, and can we relate them back to normative expectations expressed as technical fairness criteria?

The first step in this analysis is to understand if and why each formulated bias is truly harmful, i.e. identify the forms of harm caused by each type of bias.

\subsection{Bias and Harm}
As a reference taxonomy of harm, we adapt the taxonomy by Shelby et al.\cite{harm-shelby2022sociotechnical}:
\begin{itemize}
    \item \textbf{Allocative harms} occur when a algorithmic system deprives certain social groups based on disability, ethnicity, gender identity, religion or a specific characteristic, of information, opportunities or resources and allocates those resources to other dominant groups. 
    \item \textbf{Representational harms} are caused by representations of a social group in a less favorable light than others, demeans them, under-represents them or fails to recognize their existence altogether (adapted from \cite{blodgett2020language}); 
    \item \textbf{Quality-of-service harms} include disparities in the performance of algorithmic systems that do not provide the same service quality to different groups of people \cite{harm2-bird2020fairlearn}.  
    \item \textbf{Interpersonal harms} occur in four situations: When individuals lose their agency due to the use of an algorithmic system \textit{(loss of agency)} or feel that they are being controlled by an algorithmic system \textit{(social control)}; When technology is used to harass people (e.g. impersonating, cyberbullying, etc.)\textit{(Technology-facilitated violence)}; When an algorithmic system purposefully exploits user behaviors, or makes incorrect health inferences that result in emotional manipulation or distress \textit{(Diminished health and well-being)}; When a system collects users' data without explicit and informed consent, and users feel of being watched or surveilled \textit{(Privacy violation)}.
    \item \textbf{Societal harms} refers to  STSs that accelerate the scale of harm by using \textit{misinformation} (the spread of misleading information whether or not there is intention to deceive), \textit{disinformation} (false information) and \textit{malinformation} (“genuine information that is shared with the intent to harm”) \cite{malinfo-janzen2022cognitive}.
\end{itemize}

The forms of bias that we identified previously can be related to some of these harms, as follows.

\paragraph{Under-participation} While the lack of participation by women -- or any other group -- may not correspond \emph{per se} to any definition of harm, it is a symptom of other problems (phenomena reducing women's participation), and it may participate in causing other phenomena that do match definitions of harm. 
\paragraph{Under-representation} The under-representation of a group, as exemplified by the number of biographies of members of the group, fits the definition of representational harm. In addition, if Wikipedia biographies are considered to have the effect of increasing people's reputation and online visibility, then this phenomena acts as a ``service" which does not provide the same service to members of the affected group as it does to non-members: the harm then matches the definition of \emph{quality of service harm}.
\paragraph{Biased deletion processes} Biased deletion processes contribute to the under-representation of a group, its perspectives or its interests, which is a form of representational harm. However, when framed as a decision process that makes more errors for one group than for another, it can be seen as a form of quality of service harm, and a form of allocative harm because it deprives the group of opportunities for online representation. It is also important to consider the effect on a contributor of seeing the page they wrote being deleted: in this final sense, biased deletions cause a form of interpersonal harm.
\paragraph{Biased representation} Biased representations, i.e. those representing a group in a negative light or through a stereotypical lens, are the textbook definition of representational harm. In the case of the individual whose Wikipedia biography was targeted, we referred earlier to individual fairness: while fairness is the normative expectation, here the harm is a form of interpersonal harm where technology was being weaponized to target another person's reputation, and contributing the most damaging information to a Wikipedia Article can be considered \emph{malinformation}, as the primary goal is to cause harm. 
\paragraph{Over-policing} In the case that we described as ``over-policing", there is a binary decision process, which receives an edit to a Wikipedia page, and decides whether to accept it or to reject (revert) it. As one group's contributions are more closely scrutinized, which increases the risk of receiving a negative outcome (revert), this phenomenon matches the definition of allocative harm.

\subsection{Bias and decision fairness}

We can now address the second part of our research question, which we reformulate as follows: if \emph{bias} expresses a lack of fairness, can we also relate back each bias to a technical fairness criterion defined for algorithmic systems?

In order to answer this question, we must relate the observed biases to the general models used in the analysis of algorithmic systems. The main algorithmic system model that has been studied in terms of fairness is that of \emph{automated binary decisions systems}. In these decisions, inputs are associated with people, and the outputs (outcome of the decision) usually include a more desirable outcome (e.g. the person obtains a loan) and a corresponding ``negative outcome" (e.g. the loan is denied). In addition, it is usually assumed that there is a ``correct" decision for each person, and that an algorithmic system's decisions are not always correct. 

Among the types of biases described for Wikipedia, the following fit the model of binary decisions:
\begin{itemize}
    \item Under-representation of group: in particular, we can model the process leading to biographies being written and added to Wikipedia as a binary decision process. Considering that the ideal situation would be if all \emph{notable} people has Wikipedia biographies, and \emph{non-notable} people did not, there is also a \emph{correct} decision for each person (create the biography if the person is notable), which is not necessarily the one taken by the Wikipedia contributors.
    \item Biased deletion processes: Similarly to the previous case, article deletions are also a binary process, with a reference \emph{correct} decision: a biography should be deleted if the person is not notable.
    \item Over-policing: over-policing also refers to a binary decision process, where a particular edit should be accepted (if it complies with Wikipedia rules), or reverted. A person (the author of the edit) is also associated with each input of the decision, and will experience a negative outcome if the edit is reverted. 
\end{itemize}

Many technical criteria have been proposed to assess the fairness of binary decisions. In the above cases we are concerned with group fairness criteria, i.e. those that define fairness with respect to an identifiable social group. 

We list below the most prominent \emph{group fairness} criteria, and discuss their applicability to the above biases. We define the fairnesss criteria formally, using the following notations: the membership of a person to a group is represented by a random variable $A \in \{0, 1\}$, the ``correct" decision for this person is represented by a random variable $Y \in {0,1}$, and the decision output by the algorithmic system is $\hat{Y} \in {0,1}$. 

\subsubsection{Demographic Parity} One of the simplest fairness criteria, also called independence or statistical parity~\cite{tab-demographic1-zafar2017training,tab-demographic2-zhang2018mitigating}, demographic parity requires that a person's chance of obtaining the positive outcome from the decision system ($\hat{Y}=1$) is independent of the sensitive attribute: 
\begin{equation}
\label{eq:demparity}
P [\hat{Y} = 1 | A = 0] = P [\hat{Y} = 1 | A = 1]
\end{equation}

Demographic parity expresses that the members of the considered group (e.g., women) should receive a positive output at the same rate as those from the complement group (i.e. men). In the bias situations above, using demographic parity as a reference fairness criterion means ignoring the ``correctness" of decisions, i.e. whether the outcomes of the process comply with Wikipedia rules (published biographies are those of notable people, accepted edits comply with the rules, etc.). When the number of women's biographies is simply compared with that of men's biographies (i.e. only 17.9\% of biographies are women's biographies\cite{02-tripodi2021ms}), the implicit expectation is that it should be 50\%, meaning that men and women should have biographies at equal rates: in other words, the normative expectation is demographic parity. Similarly, observing that women's biographies are deleted more (or less than men's), and only comparing with the total number of biographies available to delete, is also a way of expressing that the normative expectation is demographic parity. 


\paragraph{Equal opportunity} Also called Separation and Positive Rate Parity \cite{tab-14-hardt2016equality}: this criterion defines fairness according to the distribution of predictive errors across groups: \emph{equal opportunity} means that people for whom the positive outcome is the correct decision have the same probability (across groups) of obtaining this decision from the decision system: 
\begin{equation}
    P [\hat{Y}  = 1|Y = 1, A = 0] = P[\hat{Y}  = 1|Y=1, A = 1]
\end{equation}
\paragraph{Equalized odds} The extension of \emph{equal opportunity} to all outcomes: the probability of the negative outcome being correctly attributed is also equal across groups.
\begin{equation}
    P [\hat{Y}  = 1|Y = y, A = 0] = P[\hat{Y}  = 1|Y=y, A = 1], \forall y \in {0,1}
\end{equation}

Unlike demographic parity, equalized odds does consider the \emph{correct} decision in determining whether decisions are fair towards a social group. In our context, if we consider the process of publishing Wikipedia biographies, applying equalized odds as a reference fairness criterion means verifying whether \emph{notable} women (or members of the considered group) obtain biographies at the same rate as \emph{notable} men. This assessment of bias will differ from demographic parity if the proportion of notable people is different from one group to another. As the number of men and women's biographies in Wikipedia is very unbalanced, the analysis made by Wagner et al.\cite{01-wagner2015s}, concluding that there is no bias against women, is explained by this formulation of fairness. We note that the same concept is applied by Adams et al. \cite{adams2019counts}, who nevertheless obtain different results. This can be explained by the fact that the two studies use different datasets to represent notability, which serves as the ``ground truth" in this formulation.

If we compare the two perspectives (applying demographic parity or equalized odds), in the latter case fairness is achieved if notability criteria are applied correctly, with no biases against any group. In the former case, this will not be sufficient to achieve fairness since the ``ground truth" criterion is itself biased (e.g. men achieve ``notability" (in the Wikipedia sense) at a higher rate than women), and therefore the implication is that notability criteria should be changed in order to achieve the true fairness represented by demographic parity.


\subsection{Bias and fairness in representation}
 
 The other perspective on fairness in algorithmic systems relies on the concept of ``representations", without any particular model of the system behavior. In this model, fairness is defined as the lack of \emph{representational harm} \cite{crawford2017trouble,harm-shelby2022sociotechnical}. 
 
 As noted previously, not all of the identified bias phenomena match the model of a binary decision: the remaining bias phenomenon that we have identified is \emph{biased representations of group members}. As we noted above, the harm associated with this bias is primarily \emph{representational harm}. 

 For the different Wikipedia bias phenomena that fit under this general ``biased representation" concept, we turn to criteria for measuring representational harm in text. The main applicable techniques aim to detect \emph{negative stereotypes and demeaning language} in the representation of social groups.

These biases are conveyed by the semantics of text, and in modern natural language processing (NLP) algorithmic systems, semantics are usually encoded in the system's internal language representation, i.e. a language model trained on a large corpus of naturally occurring text. The most common language modelling technique, known as \emph{word embedding}, consists in mapping words to high-dimensional vector spaces in such a way that words with similar meanings are mapped close to one another in the vector space.

Biases can be measured in word embeddings using a technique adapted from the Implicit Association Test \cite{greenwald1998measuring} used to identify biases in humans: essentially, the test measures whether a word referring to the social group is more similar in the semantic space to pleasant or unpleasant terms, which indicates whether the language model is biased positively or negatively with respect to this group~\cite{caliskan2017semantics}. In addition to capturing overall sentiment towards a group, similar measurements can capture stereotypes and societal biases such as the idea that housework is primarily a task for women~\cite{ML01-causality-bolukbasi2016man}. The technique was also extended to evaluate biases in full sentences using sentence encodings instead of individual word embeddings~\cite{may-etal-2019-measuring,tan2019assessing}: the tests for words are known as WEAT tests, and sentence encoding tests are known as SEAT tests.

\paragraph{Application to Wikipedia}
in the Wikipedia setting, representation biases are typically measured by selecting words or concepts that represent a particular stereotype, and counting their occurrences across articles associated with the different groups (e.g. the biographies of men and women, or articles on the same topic in different language editions). To measure biases in a more general way, semantic representations could be derived from Wikipedia (in fact, several existing semantic models are trained on corpora including the English Wikipedia), then evaluated for bias using the above techniques (SEAT and WEAT). The difficulty in these experiments would be to re-train language models on a subset of Wikipedia articles to be evaluated: this may not be possible in the case of small or even moderate-size datasets, because language models require large amounts of training data. 

\section{Causality relationships}
\label{sec:causal}
Having explained the observed biased in terms of sociotechnical harm and fairness, we now turn to the causes of these different phenomena. We consider our third research question: 

\textbf{(RQ3)} Are there any clear causal relationships between these problematic phenomena?

The purpose of answering this question is to identify the causal structure of the situation: how do the different problematic phenomena observed in the complex socio-technical system influence one-another? How might we affect the overall situation if we attempt to mitigate one of the phenomena?

Again, we consider each phenomenon (table \ref{tab:paper-list2}) as an abstract type of bias.
\subsection{Under-participation}
\paragraph{Phenomenon} members of a group participate less in Wikipedia than members of other groups.

\paragraph{Causes} in the case of women, a fairly wide variety of causes has been discussed to explain their low participation in Wikipedia. In particular, Sue Gardner, a former executive director of Wikipedia, identified in 2011 nine reasons for this\cite{gardner2011nine}, most of which are discussed by other papers reviewed for the present study. These reasons can be summarized by the following three general causes: (i) self-exclusion, (ii) the unwelcoming atmosphere of Wikipedia, and (iii) the poor design of Wikipedia's interface. 

The first, self-exclusion, covers purely social causes why women cannot, or choose not to participate in Wikipedia. By ``purely social" we mean that these causes are unrelated to the design or atmosphere of Wikipedia itself: women's lower of self-confidence compared to men, their already busy lives (jobs and a disproportionate share of house work). 

The second general cause, which we refer to as the unwelcoming atmosphere of Wikipedia, refers to social factors specifically in how they occur within the community of editors: the conflictual culture \cite{EditingWarsBehindtheScenes,03-ratkovic2021women}, a sometimes misogynist and/or ``sexual" atmosphere \cite{gardner2011nine}. 

Finally, the third cause is linked to the more technical aspects of Wikipedia, including its graphic user interface and the expected interaction protocols with other members of the community. An important aspect of this general cause is that it is the only one that can be directly addressed by any technical changes to the Wikipedia STS itself. 

The causes of other groups' under-participation in Wikipedia have not been well studied. We conjecture that they are related to similar factors: self-exclusion factors (e.g. language barrier), unwelcoming atmosphere of Wikipedia (if the existing community of contributors is predominantly from a group with very different social norms from their own), and possibly user interface factors. 

In addition to these factors, two additional factors should be added, related to other bias phenomena on Wikipedia. First, the problem of ``over-policing" clearly leads to lower participation of members of a group affected by this phenomenon. As we do not know the demographic make-up of anonymous users (e.g. in terms of gender) we cannot infer how over-policing of anonymous users may affect the participation of groups defined by other attributes (e.g. women or visible minorities), but there is a clear possibility that a causal relationship occurs here. Secondly, we would conjecture that the under-representation of a group's interests in Wikipedia would also affect the group's participation in Wikipedia, by reinforcing the idea that Wikipedia is not ``meant for them".

\subsection{Under-representation of group members}
\paragraph{Phenomenon} members of a group are under-represented in the content of Wikipedia articles, either in terms of biographies or by mentions in other articles.

\paragraph{Causes} the under-representation of women in Wikipedia is related to three causes: one is the under-participation of women: assuming that women are more likely to be more interested than men in writing about other women, it stands to reason that a smaller community of women contributors will create fewer biographies or women or mentions of women in other articles, as compared to the hypothetical situation where the pool of contributors was more balanced. Similar phenomena most likely occur for other social groups, such as citizens of a given country or members of a tightly-knit cultural group: we would expect member of the group would be more likely than others to write about their fellow members. A second relevant reason is the influence of the \emph{notability} rules on Wikipedia. As indicated by the debate on whether women are actually under- or over-represented among notable people (\cite{01-wagner2015s,adams2019counts}), Wikipedia's notability rules define (vaguely) who is allowed a Wikipedia biography. These notability rules will also combine with Wikipedia's deletion processes to remove some of the existing biographies, and thus also indirectly affect the representation of groups. 

Finally, we note that another indirect causality relationship may exist between the under-participation of group members and their under-representation: the more active participants of Wikipedia contribute to shape Wikipedia norms and policies, including notability rules. A group for whom ``external" notability 
 criteria (e.g. references in print newspapers from a certain country) is biased may recognize that relying on this notability criteria is a bad idea, and push for change. We therefore conjecture the existence of an additional causal connection between the under-participation of group members and the biases of notability rules towards that group.

\subsection{Under-representation of a group's interests and perspectives}
\paragraph{Phenomenon} few articles discussing topics of particular interest to a group, or the perspectives of the group being under-represented in articles in general. 
\paragraph{Causes} as is the case with under-representation of a group, the under-participation of the group would also cause an under-representation of the group's interests and perspectives, in the same way. Generally speaking, societal biases also find their way into the contents of Wikipedia, and contribute to silencing the perspectives of other groups: here, the effect is that members of other groups, instead of merely failing to include the group's perspective in and article, may actively suppress it~\cite{colonization-morris2022}.

\subsection{Biased representation of a group}
\paragraph{Phenomenon} the content of article conveys biased representation of members of particular social groups. 
\paragraph{Causes} in this case, the dominant reason for these biases is the societal biases present in the greater society: crowdsourced text naturally mirrors the biases in the minds of its authors. However, one may expect that members of the affected group would want to correct this situation, by editing  the biased articles in order to reduce their bias; in this sense, the lack of participation from a group is also likely to contribute to biased representations of the group. 

\subsection{Over-policing bias}
\paragraph{Phenomenon} the edits from a particular group (anonymous users) are more closely scrutinized than those of other groups (registered users).

\paragraph{Causes} according to De Laat\cite{edit-flagging2de2016profiling}, this phenomenon occurs when editors patrol the new edits of Wikipedia and misinterpret the interface of automated flagging tools: as they are aware the anonymous groups are more likely to vandalise articles or produce ``bad" contributions (according to Wikipedia policies), they pay extra attention to edits produced by anonymous users, rather than simply relying on the flagging software, which already considers the anonymity of users as a risk factor.

\subsection{Summary}
\begin{figure*}[tb]
\begin{center}
  \graphicspath{ {./images/} }
  \includegraphics[scale=0.50]{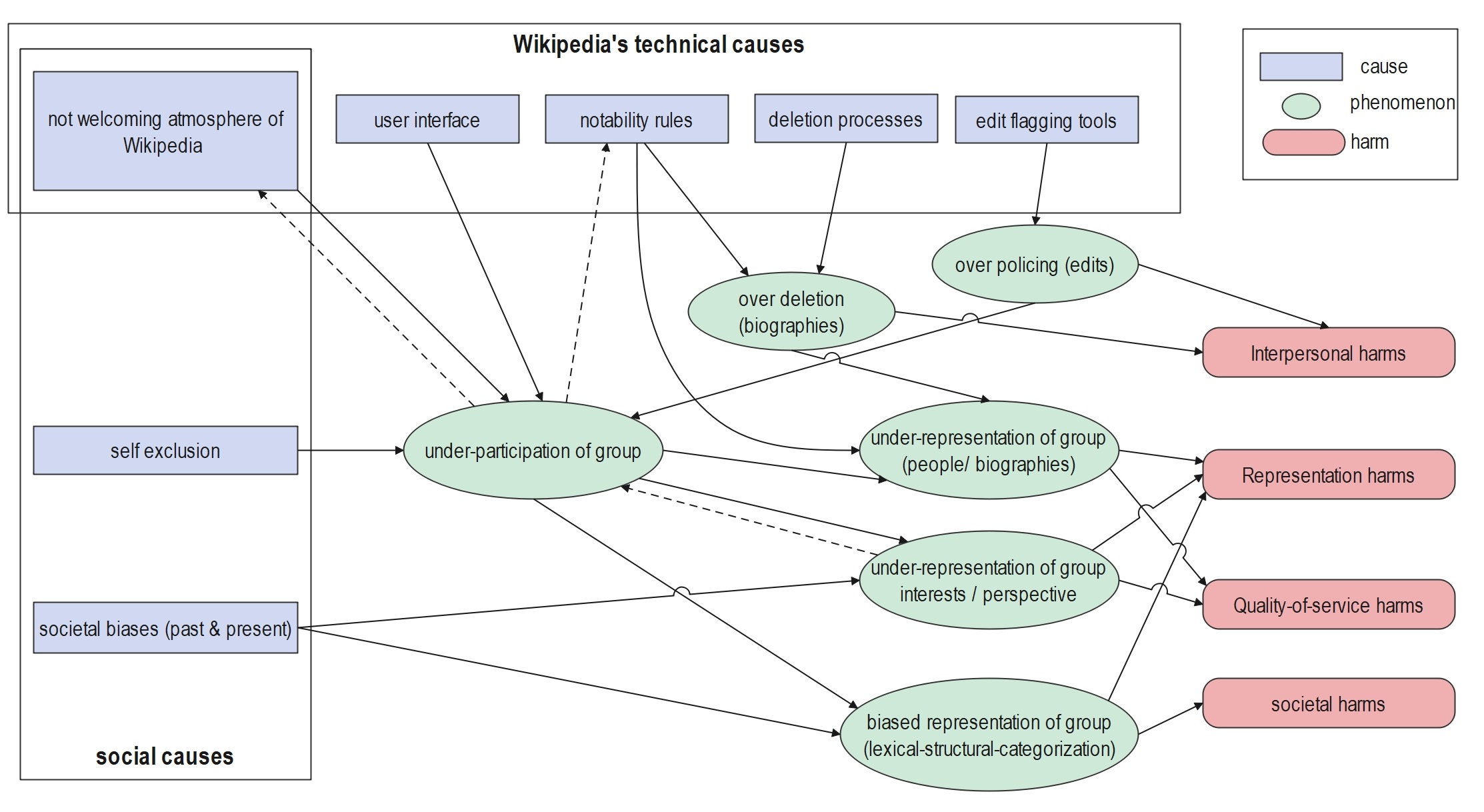}
  \caption{Causality relationship network for Wikipedia's problematic phenomena.}
  \label{fig:causality}
  \end{center}
\end{figure*}

The different causal relationships discussed in the previous subsections are illustrated in figure \ref{fig:causality}.

The causes are marked with blue rectangles, and are depicted in two groups: (i) social causes (that not directly related to Wikipedia's technical part) on the left side of the figure and (ii) Wikipedia's technical causes which include all of the processes, rules, and technical tools (including bots of wikipedia) on the top of the figure. The cause in the intersection of social causes and Wikipedia's technical causes in the top left, (not welcoming atmosphere of Wikipedia) as it is both a social cause (it is directly associated with the attitudes of people) and a technical cause, in the sense of being within the scope of Wikipedia as an STS.

Bias phenomena are shown with green ellipses, and harms as red rounded rectangles. Arrows indicate causal relationships documented in the literature; dashed arrows are additional causal relationships which we did not see explicitly formulated in the literature, but that we believe to be true. Verifying these empirically is part of future work. 


\section{Conclusion}
\label{sec:conclusion}
In the perspective of auditing socio-technical for fairness, we present a case study of Wikipedia, where we obtain an overview of all the relevant fairness-related issues through a systematic review of 75 research papers on the topic. We identify and categorize the different phenomena associated with bias, normatively framing them as harmful with specific meanings, and in reference to specific fairness concepts first defined for algorithmic systems. 
We also describe the network of causal relationships between different problematic phenomena, and external social factors.

One of our key findings is that this STS exhibits many biases, but no clear \emph{algorithmic} biases (biases observable as the output of algorithms). There are different problematic processes, some of which are partially automated (or supported by software tools with no demonstrated biases, e.g. tools to rank potential ``bad edits"). 

The lack of purely algorithmic biases and the network of causal relationships between problematic phenomena demonstrates the need for more holistic auditing practices for STS: if there is any hope of mitigating biases in an STS such as Wikipedia -- where there is no specifically harmful algorithm to fix, and no biased dataset to unbias -- then we must focus on the socio-technical processes. These are also algorithms, in a "social computing" sense, and an important question is how interventions on this type of algorithms can help with different types of bias.

\end{document}